# Participant: A New Concept for Optimally Assisting the Elder People

Hong Sun, Vincenzo De Florio, Ning Gui, Chris Blondia
*University of Antwerp*
*Department of Mathematics and Computer Science, PATS group, Antwerp, Belgium*
*and Interdisciplinary institute for BroadBand Technology, Ghent-Ledeberg, Belgium*
*{hong.sun, vincenzo.deflorio, ning.gui, chris.blondia}@ua.ac.be*

*Abstract*

*Elder people are becoming a predominant aspect of our societies. As such, solutions both efficacious and cost-effective need to be sought. The approach pursued so far to solve this problem used to increase the number of people working in the health sector, e.g. doctors, nurses, etc. This increases the costs, which is becoming a big burden for countries.*

*In this paper we propose a new concept in the health management of elder people, which we name as "participant". We propose the "participant" concept to encourage elder people to participate in those group activities that they are able to. Their roles in these activities are not passively requesting help, but actively participating to some healthcare processes. Characteristics of the participant approach are that medical resources are efficiently spared with this model, and the social network of the elder people is kept. A "virtual community" for mutual assistance is set up in this paper, and the simulations demonstrate that the "participant" model could fully utilize the community resources. Furthermore, the psychological health of the elder people will be improved.*

## 1. Introduction

As well known, the proportion of elderly people keeps increasing since the end of last century. The European overview report of Ambient Assisted Living (AAL) investigated this trend [1]. Studies of Counsel and Care in UK found out that these elderly people would prefer to live in their own house rather than in hospitals, thus they need support to remain independent at their home [2]. In order to improve the quality of life for the elderly and disabled people, it is important to guarantee that assistance to those people be timely arranged in case of need.

Assistive devices are developed to facilitate the daily lives of these elderly and disabled people. But assistive devices also have their limitations: For instance, in the AAL country report of Finland, it was remarked that "the (assistive) devices are not useful if not combined with services and formal or informal support and help" [3]. We share this view and deem informal carers as indispensable when constructing timely and cost-effectively services to assist the elderly people. We developed a design tool to evaluate the performance of informal carers in so-called mutual assistance communities [4], i.e. communities whose members may request assistance and at the same time get motivated to play the role of caregiver. Simulations have shown that informal carers are indeed capable to contribute effectively to the community welfare.

In this paper, we continue along that track and propose a new concept for the health management of elder people, which we name as "participant". Traditional care system are mostly focused on maintaining the physical health of the elder people, treating them as someone highly dependant on service from outside world, which in fact frustrates their mental health. In our "participant" model, elder people will be encouraged to participate in the activities they are able to. Their roles in these activities are not passively requesting help, but actively participating. The model not only takes care of the health of the elder people, but also their happiness, bringing them the chance to participate in the activities which can not be obtained from pure medical care. Their social network will be maintained, physical practice will be encouraged to be taken, and a number of their needs will be met by activating or participating in a group activity, rather than calling service from professional carer, so that the social-medical resources are saved.

The remainder of the paper is organized as follows: In Section 2, related work will be reviewed. In Section 3, an overview of the participant model is presented. Simulations are shown in Section 4, and conclusions and future work are given in Section 6.

## 2. Related work

Numerous researches are being carried out on building intelligent environments around people, such as Aware Home [5], Smart In-Home [6], BelAmI [7], I-Living [8]. These researches on "smart houses" improved the independency of the elderly people, and reduced the required manual work. Their limitation is the lack of communication with the community outside the house, which inherently limits the service exploration, and may isolate the user from the outside world.

Some projects begin to focus on the communication with the outside community. One such project is COPLINTHO [9], which built an eHomeCare system combing forces from the patient's family, friends and overall care team. The limitation of this class of investigations is that the application is restricted to the recovery progress of a patient, thus the communications are mainly focused on exchanging the medical data of the patient. Finally, some efforts aim to assist the old people by building a community. One such case could be found in Beer's work [10] [11]. But no mechanism is foreseen to locate and map services, and his system only scheduled the professional carer.

All these above represent examples of the existing approaches to AAL, which we may classify as person-oriented, family-oriented, and community-oriented. Their achievements have both inspired our work and provided us with useful provisions, methods, and architectures that integrate our contribution, but none of these methods is a true combination of technical and social forces. We intend to build up a mutual assistive community, where the dwellers could interactively communicate, provide informal care and jointly participate in group activities.

In mutual assistance communities, service requesters and providers, even from the same ontology domain, are heterogeneous in nature and capabilities. The effectiveness of the participant scheme relies on the ability of the service center to understand service requests and locate, rank and deploy candidate participants. Our research on constructing the service center for semantic reasoning and matching can be found in [12]. In this paper, we focused on investigating how the participant model could save the medical care resources of a community.

## 3. Participant model

### 3.1. The concept of participant

The concept of participant carries on the concept of informal carer. We envisage that some activities which the elderly people want to engage in may need more than one people to participate, such as walking in the park with someone else, playing chess, etc. The conventional approach to fulfill these requests would be, e.g. to send a nurse to the requester. With the assistance of informal carer, before requesting the nurse service, someone living nearby would be sought and, if successfully identified, would be asked to provide help. The introduction of the informal carer could save the professional medical resources. But the elder people are always passively receiving help and losing their independence in both situations.

The concept of participant provides a third solution to meet these requirements of the elderly people. When elderly people want to initiate or join a group activity, they will send a request to participate this activity. The request will be parsed by a service center. If such a participation is ongoing, the requester could join this activity directly, otherwise, based on the time constraint of the requester, the system will either initiate a new joint activity or try to find service from informal or professional care-givers to fulfill the user's requirement.

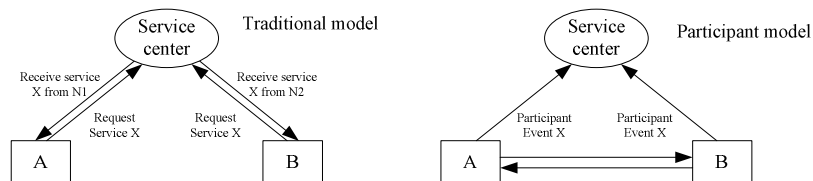

**Figure 1. Comparison between traditional model and participant model**

Figure 1 compares the participant model with the traditional one. A and B represent elder people, N1 and N2 represent care-givers. When A and B want to participate a same event, the service center will try to establish a link between them in the participant model rather than requesting for help in the tradition model. This not only decreases social costs, but also encourages social contacts and produces self-esteem.

### 3.2. A practical model

Figure 2 shows the process to parse a requirement in our proposed mutual assistance community. The requirement from a requester is divided into three categories: alarm signal, non-urgent request and activity participation request. The alarm signal may be triggered by the user or assistive devices, it has the highest priority. The participant signal represents the people's willingness to participate an activity, and has the second priority. The non-urgent request for care is left with the lowest priority.

The requirement of an elder person could be fulfilled by the following three players:

**Professional Carers (PC):** Those who have specialized intellectual or creative expertise based on personal skills, education and experience. **Informal Carers (IC)**: This type of service providers normally are ordinary people, volunteers or specifically employed people. **Participants**: Those who are willing to join certain activities. They either join an established group activity, or initiate one.

The process of parsing a user's request with the "participant" approach is as follows:

1. Once a requirement signal is received, check whether it is an alarm, if yes, turn to step 6, if no, then turn to step 2.

2. Check whether the requirement signal is a participant signal, if no, turn to step 3, otherwise, turn to 2.1.

2.1 Check whether there is ongoing participation activity as required; if so, join this activity, otherwise, turn to 2.2.

2.2 A new participation activity will be initialized as the user required.

2.3 Check whether the time constraint is met; if so, turn to step 3, else, keep "wait".

3. The requirement will be treated as a non-urgent request.

4. The service center will look for a carer for the requester based on the requester's preference. This can either be a professional carer or an informal carer.

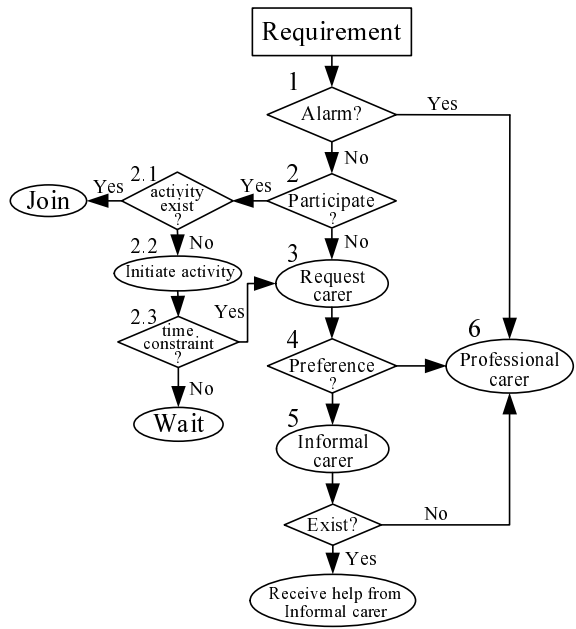

Figure 2. Process to parse a request

5. The requester prefers informal care. If an informal carer could be found, the requester could receive help from that informal carer, otherwise, turn to step 6.

6. The requester prefers professional care. Professional care is assumed to be available and able to fulfill any kind of request.

## 4. Simulation

The system proposed in this paper is modeled by a 25 by 25 cell grid; each cell in this grid represents an individual, and the whole grid represents a community. All the cells in the simulation are constructed homogenously with a same prototype. According to the predefined rates, every cell is randomly assigned to different states in the initialization stage. Neutral cells will randomly change their states after initialization according to predefined probabilities. The experiments will investigate how the system behaves with different rates of different cells.

In the simulation, with every specified setting, we run the system for 10000 steps after initialization. The status of the grid is recorded every 10 steps, thus 1000 results are produced for every setting. The average of these 1000 results will be used to reflect the system performance under different settings.

| PCrate: | 0.1  | IC_d: | 0.05 |
|---------|------|-------|------|
| ICrate: | 0.25 | N_d:  | 0.05 |
| Rrate:  | 0.25 | A_d:  | 0.15 |
| Nrate:  | 0.4  | P_d:  | 0.15 |
| A rate: | 0.15 | R_d:  | 0.7  |
| P rate: | 0.35 |       |      |
| R rate: | 0.5  |       |      |

| PC #:   | 24.0; 14.876  |
|---------|---------------|
| IC #:   | 47.764; 5.091 |
| Ne #    | 92.359; 92.359|
| A #:    | 8.281; 0.0    |
| P #     | 8.364; 1.504  |
| R #     | 44.232; 5.015 |
| Failure | 384           |
| Latency | 23.346        |

Figure 3. Screenshot of one setting            Figure 4. Corresponding result

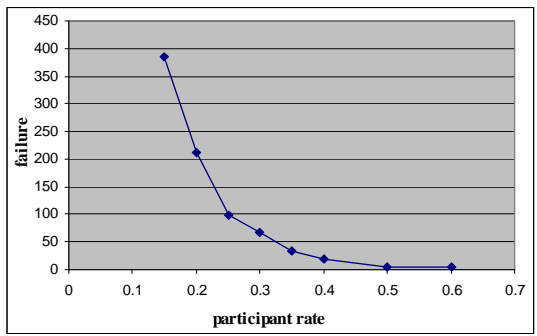 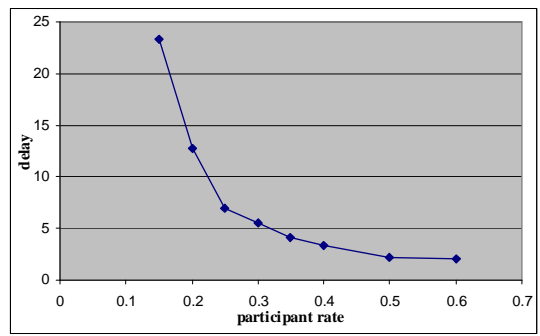

**Figure 5. Experienced failures**   **Figure 6. Experienced delays**

Figure 3 shows the screenshot of one simulation setting. The setting shown in Fig. 3 is a very stringent situation: In the beginning, the rate of PC, IC, Requester, and Neutral are 10%, 25%, 25%, and 40% respectively. Among the Requesters, alarm requests count for 15%, participant requests count for 35%, and non-urgent requests count for 50%. After initialization, there will be 5 cells randomly picked to randomly change their cell types (the selection excludes the PC and cells who are actively interacting). The selected cell will have 5% probability to be an IC, 5% to be Neutral, and 90% to be a Requester. Clearly this is a challenging situation as the probability to be a requester is rather high. When the cell is selected to be a Requester, the probability to have an alarm request is 15%, that of a participating request is 15%, and that of a non-urgent request is 70%.

Figure 4 shows the corresponding status of the community, based on averaged values, at the end of the simulation run. The indexes labeled with "#" show the averaged number of PC, IC, etc. In each field, the former number indicates the average number of people, while the latter one means the number of cells which are not active. For example, after a simulation run on the average the community hosts 47.764 informal carers, 42.673 of which are providing help while 5.091 are not interacting (that is, they are available). The number of failures indicates the total number of aborted requests accumulated in 1000 results; this is not an averaged value. Field "ave latency" indicates the averaged steps that all the requester cells had to wait in every step.

The amount of failures and delays experienced is shown in Fig. 5 and Fig. 6 respectively. The setting of the system follows the values in Fig. 3 except for the participant rate and R rate change. The participant rate here indicates the probability to send a participating request if a cell is selected to change its type into a requester, and the R rate represents the probability to send a non-urgent request.

Main conclusion of our simulations is that both the failure numbers and the latency decrease with the increase of participant rate. When the participant rate is 15%, there are 309 failures, and the average latency is 16.3 time steps. When the participant rate increases to 25%, the failures drop to just 99, and the averaged latency drops to 7.0. When the participant rate reaches 60%, the failures further drop to 5 and the averaged latency drops to 2.0.

These two figures prove that the introduction of participant can greatly reduce the request failures and service delay. When the participant rate is low, its impaction is very significant: a tiny increase on the participant rate will bring great improvement. What was not shown in the above figures is that when keeping the other settings but decreasing the participant rate to 10%, failures increase to 987, and the latency will reach 4864.8, which means the community becomes unstable and chaotic.

## 5. Conclusions and future work

The constant increase of the population of elder people poses enormous economic and social challenges to our society. Compared with the dedicated efforts on developing new assistive devices and technologies to help the elder people live longer at home, there are few researches focused on maintaining the mental health of the elder people. This means neglecting the fact that the loss of social network, the lack of physical activities, and the sedentary status will greatly reduce the confidence of the elder people, increasing the feeling of loneliness and frustration and reducing their living standard. Moreover, their immobility will also impact on their physical health.

This paper addresses this situation by proposing a "participant" model for mutual assistance community to meet the above mentioned challenges. We use this "participant" model to encourage the elder people participate in the group activities, thus maintaining and even establishing new social networks. Through that, elders can also improve their physical health by taking physical activities. Moreover, when these elder people are actively joining the group activities, part of their daily requirements are expected to be solved by those group activities. Thus elder people can become less dependent on professional or informal medical care. Consequently, social and medical resources could be saved greatly.

A practical participant model was proposed in this paper. We simulated this model, and the simulation results support that the introduction of participant model may greatly reduce the service delay and request failures. Especially when the medical resources are impoverished, the introduction of "participant" model could fully utilize the community's resources and make best efforts to meet the dweller's requests.

Further work will focus on the integration of this "participant" model with our current research on service-oriented infrastructures for ambient assisted living [12]. By modeling the service description as Semantic Web Services, various requests can automatically be discovered, reasoned about and mapped. Various types of participant requests could be categorized and mapped in an automatic and efficient way.

## 6. References

[1] Horst Steg, et al. Ambient Assisted Living – European overview report, September, 2005.

[2] Counsel and Care, Community Care Assessment and Services, April, 2005.

[3] Ambient Assisted Living, country report, Finland, 2005.

[4] H. Sun, V. De Florio and C. Blondia. A design tool to reason about Ambient Assisted Living Systems. Proceedings of the International Conference on Intelligent Systems Design and Applications, 2006.

[5] Georgia Institute of Technology, Aware Home, http://www.cc.gatech.edu/fce/ahri/, 2006.

[6] University of Virginia, Smart In-Home Monitoring System, http://marc.med.virginia.edu/projects_smarthomemonitor.html

[7] Bilateral German-Hungarian Project on Ambient Intelligent Systems, http://www.belami-project.org/

[8] University of Illinois at Urbana-Champaign, Assisted Living Project. http://lion.cs.uiuc.edu/assistedliving

[9] Interdisciplinary institute for BroadBand Technology, COPLINTHO, Innovative Communication Platform for Interactive eHomeCare. https://projects.ibbt.be/coplintho/, 2006

[10] M. Beer, W. Huang and A. Sixsmith. Using agents to build a practical implementation of the INCA (intelligent community alarm) system, Intelligent agents and their applications, Heidelberg, Germany, 2002.

[11] M. Beer and R. Hill. Using Multi-Agent Systems to Manage Community Care. Proc. of the 10th International Conference on Knowledge-Based & Intelligent Information & Engineering Systems, 2006.

[12] N. Gui, H. Sun, V. De Florio, and C. Blondia. A Service-oriented Infrastructure Approach for Mutual Assistance Communities, to appear in the Proc. of the First IEEE WoWMoM Workshop on Adaptive and DependAble Mission- and bUsiness-critical mobile Systems (ADAMUS 2007), 2007.